\documentstyle[cjaa209,twoside,epsf]{article}
\input{psfig.sty}                               

\begin{document}

\title{What SWIFT has taught us about X-ray flashes and long-duration 
gamma-ray bursts}

\author{A. De R\'ujula$^{1,2}$
\mailto{}
}

\inst{
$^1$ CERN. 1211 Geneva 23, Switzerland\\
$^2$ The Physics Department. Boston University, Boston, Mass., USA
}

\email{alvaro.derujula@cern.ch}

\markboth{Alvaro De R\'ujula}
{What SWIFT has taught us about XRFs and GRBs}

\pagestyle{myheadings}

\date{Received~~2007~~~~~~~~~~~~~~~ ; accepted~~2007~~~~~~~~~~~~~~ }

\baselineskip=12pt

\begin{abstract}
\baselineskip=12pt

Recent data gathered and triggered by the SWIFT satellite have greatly 
improved our knowledge of long-duration gamma ray bursts (GRBs) and X-ray 
flashes (XRFs). This is particularly the case for the X-ray data at all 
times, and for UV and optical data at very early times.
I show that the optical and X-ray observations  are in excellent 
agreement with the predictions of the ``cannonball" model of GRBs and 
XRFs. Elementary physics and just two mechanisms 
underlie these predictions: inverse Compton scattering and 
synchrotron radiation, generally
dominant at early and late times, respectively. 
I put this result in its proper context and dedicate the paper
to those who planed, built and operate SWIFT, a true flying jewel.

   \keywords{GRB, XRF, Supernova, Compton Scattering, Synchrotron Radiation}
\end{abstract}

%
%



\section{Prolegomena} 
\label{sect:intro}

With predestined celerity, the observations made or triggered by the 
SWIFT satellite have resulted in a wealth of  
information on Gamma Ray Bursts (GRBs), in particular on the early X-ray 
and optical afterglows (AGs) of X-ray flashes (XRFs) and
long-duration GRBs. The data
pose severe problems to the `standard' Fireball models, 
whose `microphysics' (see, e.g.~Panaitescu et 
al.~2006), reliance on shocks (see, e.g.~Kumar et al.~2007), and 
correlations based on the `jet-opening angle' (see, e.g.~Sato et al.~2007;
Burrows \& Racusin~2007),  may have to be abandoned.

\begin{figure}[]
\centering
\vbox{
\hbox{
 \psfig{file=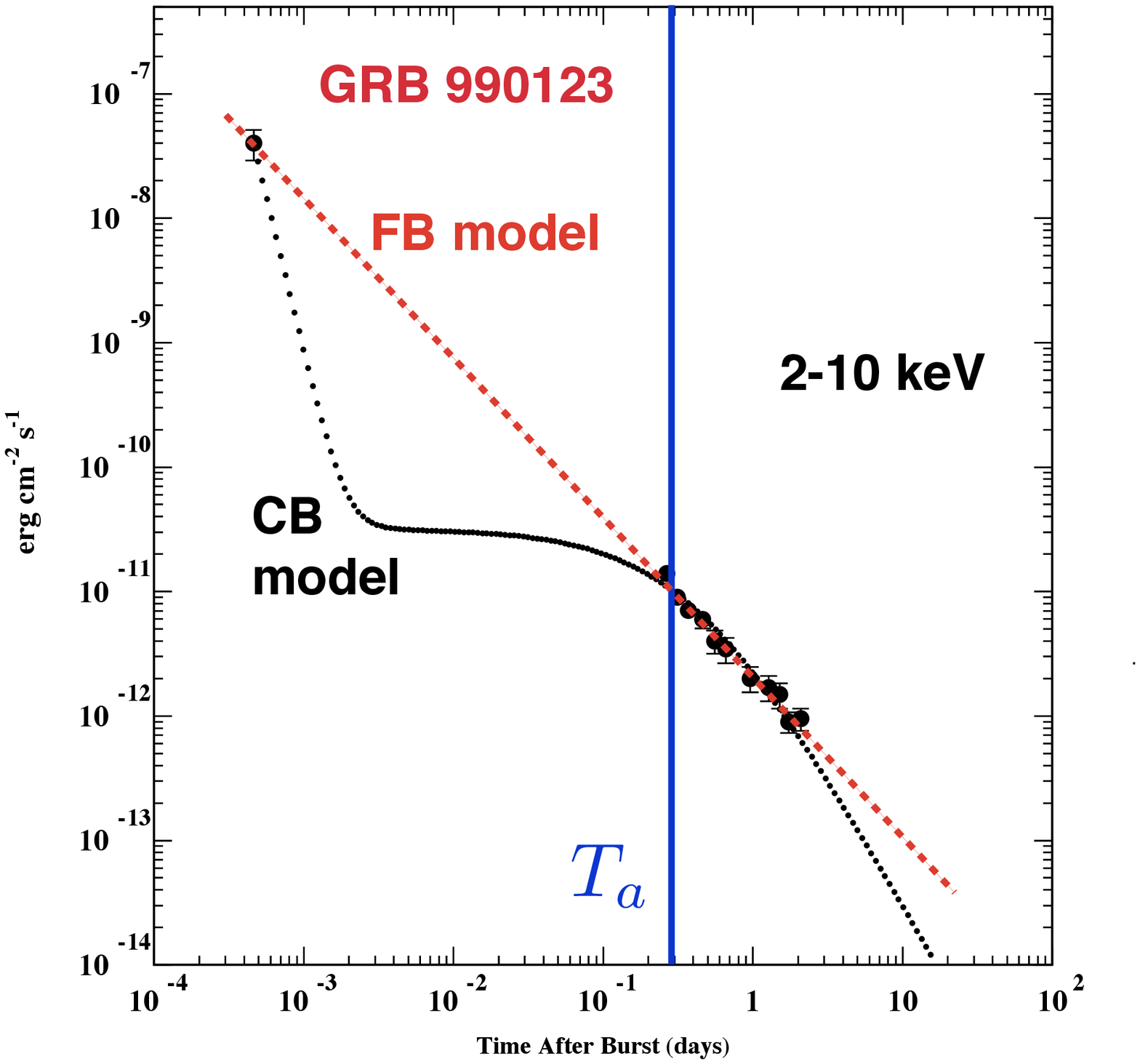,width=7.5cm}
 \psfig{file=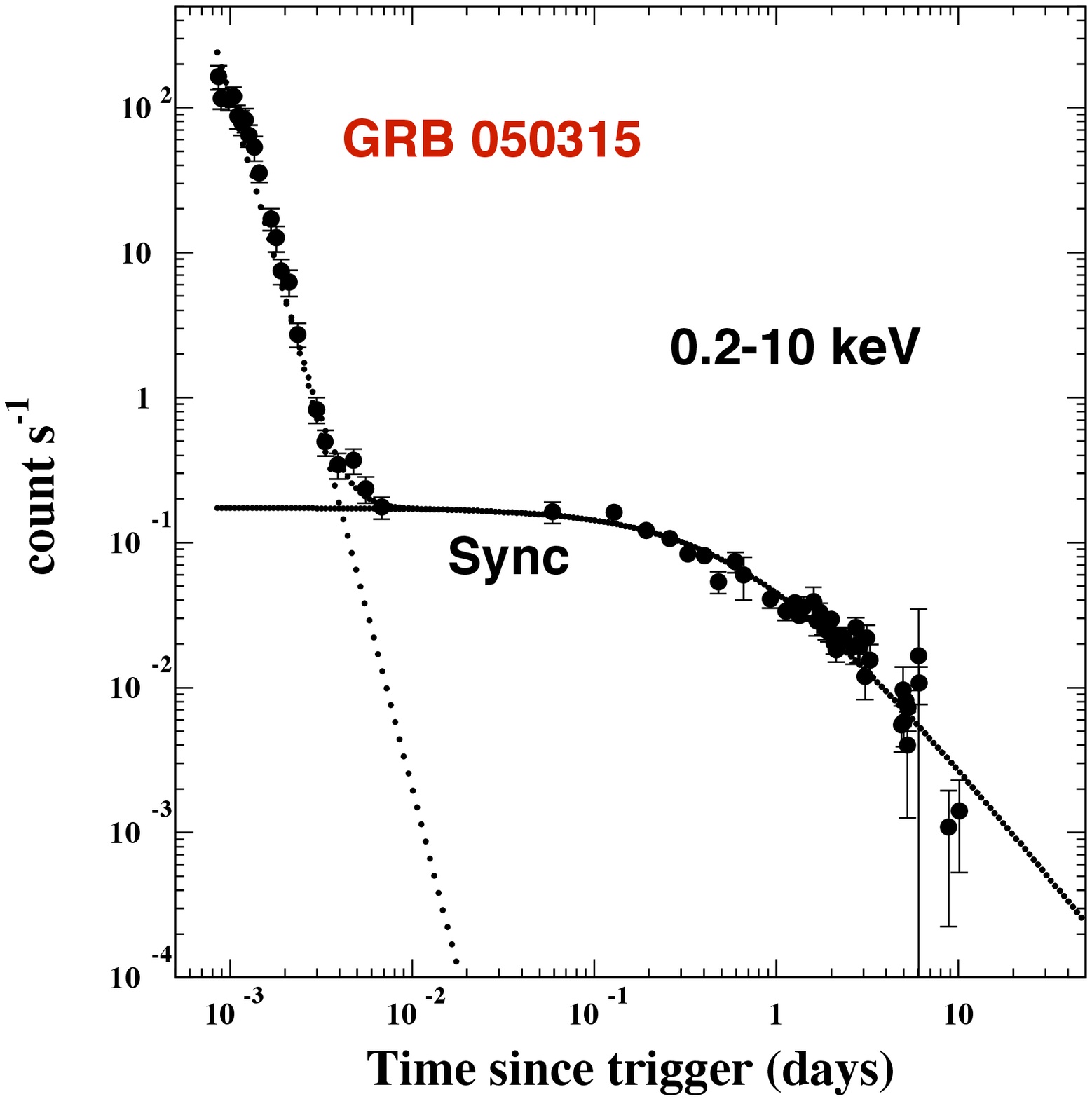,width=7.5cm }
}}
\centering
\vbox{
\hbox{
\psfig{file=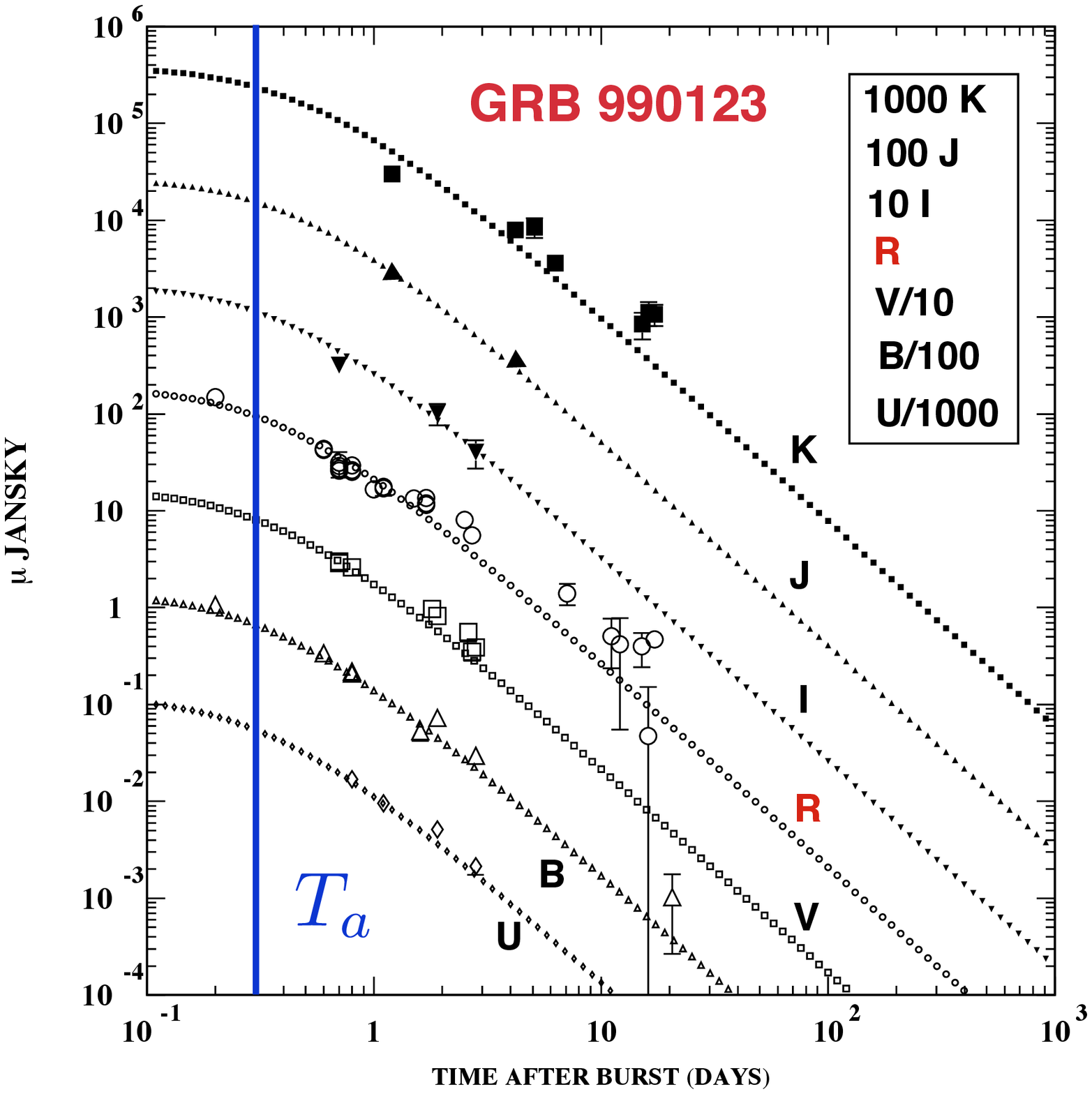,width=7.5cm}
\psfig{file=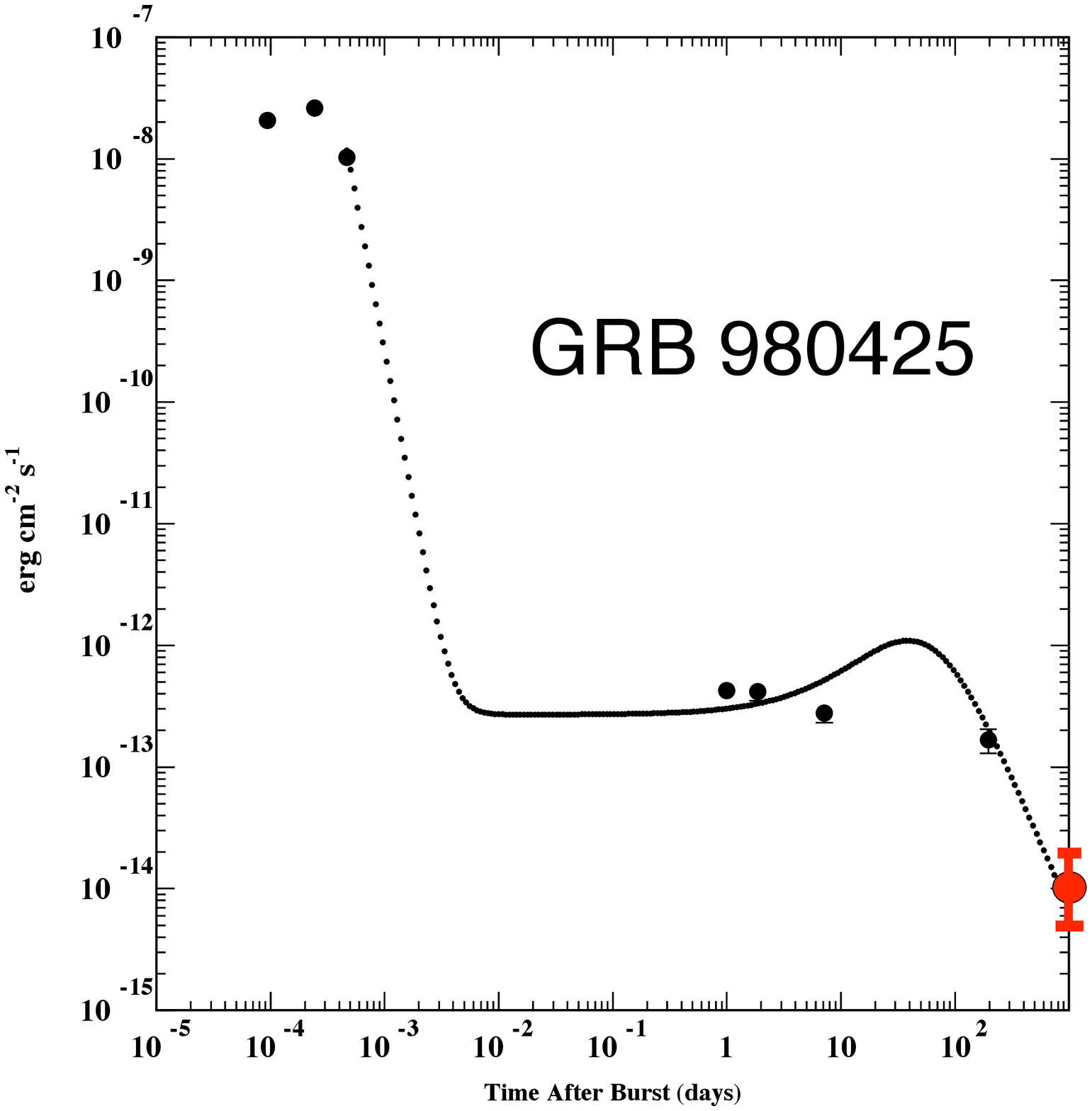,width=7.5cm}
}}
\vspace*{8pt}
\caption{{\bf (a) Top left}: Pre-SWIFT predictions for the 2-10 keV X-ray AG
in the CB (DDD02) and fireball (e.g.~Maiorano et al.~2005) models, compared to
data for GRB 990123. $T_a$ is the time ending the `plateau' phase.
{\bf (b) Top right}: Comparison between the CB model prediction and the 
canonical 0.2-10 keV X-ray light curve of GRB  050315 (Vaughan et al.~2006).
{\bf (c) Bottom left}: 
Broad band optical data on GRB 990123, fit in the CB model
(DDD03). After the onset of the plateau, the predicted and observed
evolution is achromatic from optical to X-ray energies,
as seen in this figure and its comparison with {\bf (a)}.
{\bf (d) Bottom right}: The X-ray light curve of GRB 980425
(DDD02) showing a `canonical' behaviour and what we called
a `plateau' (long and hilly in this case, given the
large observer's angle). The last (red) point postdates the original figure.
The other properties of this GRB and its 
SN are also not exceptional, in the CB model (DDD03).}
\label{f1}
\end{figure}

The said recent observations agree remarkably well with the 
predictions of the `Cannon Ball' (CB) model (Dar \& De R\'ujula~2004;
Dado, Dar \& De R\'ujula~2002;~2003, hereafter
DD04; DDD02; DDD03, respectively). Some examples are given in Fig.~\ref{f1}. 
The predicted light-curve of the X-ray AG afterglow is shown in 
Fig.~\ref{f1}a for the Fireball (see e.g.~Maiorano et al.~2005) and CB 
(DDD02) models. Many 
X-ray AGs have a `canonical behaviour' (e.g.~Nousek et 
al.~2006; Zhang~2006; O'brein et al.~2006), in impressive 
agreement with the CB-model predictions, as in the example of 
Fig.~\ref{f1}b. The evolution of the AG around the time, $T_a$, ending the 
`plateau phase' is predicted to be achromatic in the optical to X-ray 
range (DDD02),  see and compare Figs.~\ref{f1}a,c (sometimes $T_a$
precedes the end of the fast prompt decline
and no plateau is seen).
Examining SWIFT data, Curran et al.~(2006) 
found that `X-ray and optical AGs demonstrate achromatic breaks at 
about one day which differ significantly from the usual jet break in the 
[Fireball]
blastwave model of AGs'. In our model these observed `breaks'
are `deceleration bends': 
As a CB collides with interstellar matter (ISM), its Lorentz factor, 
$\gamma(t)$, typically begins to diminish significantly after
one observer's day and, consequently, so does its fluence (DDD02).
The `plateau' is nowadays
generally interpreted as a `continued activity of the engine',
yet a new surprise. In the CB model it is just the 
result of the initial, almost inertial motion of the CBs:
inevitable, not surprising. Even our closest neighbour to date,
GRB980425, associated with SN1998bw, is canonical, not exceptional.
Its X-ray light curve (DDD02, DDD03) is shown in 
Fig.~\ref{f1}d; its last data point (Pian 2002, Kouveliotou 2002)
was predicted, not fit.

In the CB model, as in the observations, there is no clear distinction
between prompt and {\it after}-glow signals. But there are two basically
identical radiation mechanisms: inverse Compton scattering (ICS) and 
synchrotron radiation (SR).
For the scores of cases we have studied, the
$\gamma$ and X-ray emissions are dominated by ICS
until the end of their fast-declining phase. 
The X-ray production, from  the `plateau' onwards,
is dominated by SR. All other
general statements equating the two mechanisms to the two phases
(prompt and afterglow)
have exceptions.

\begin{figure}[]
\centering
 \psfig{file=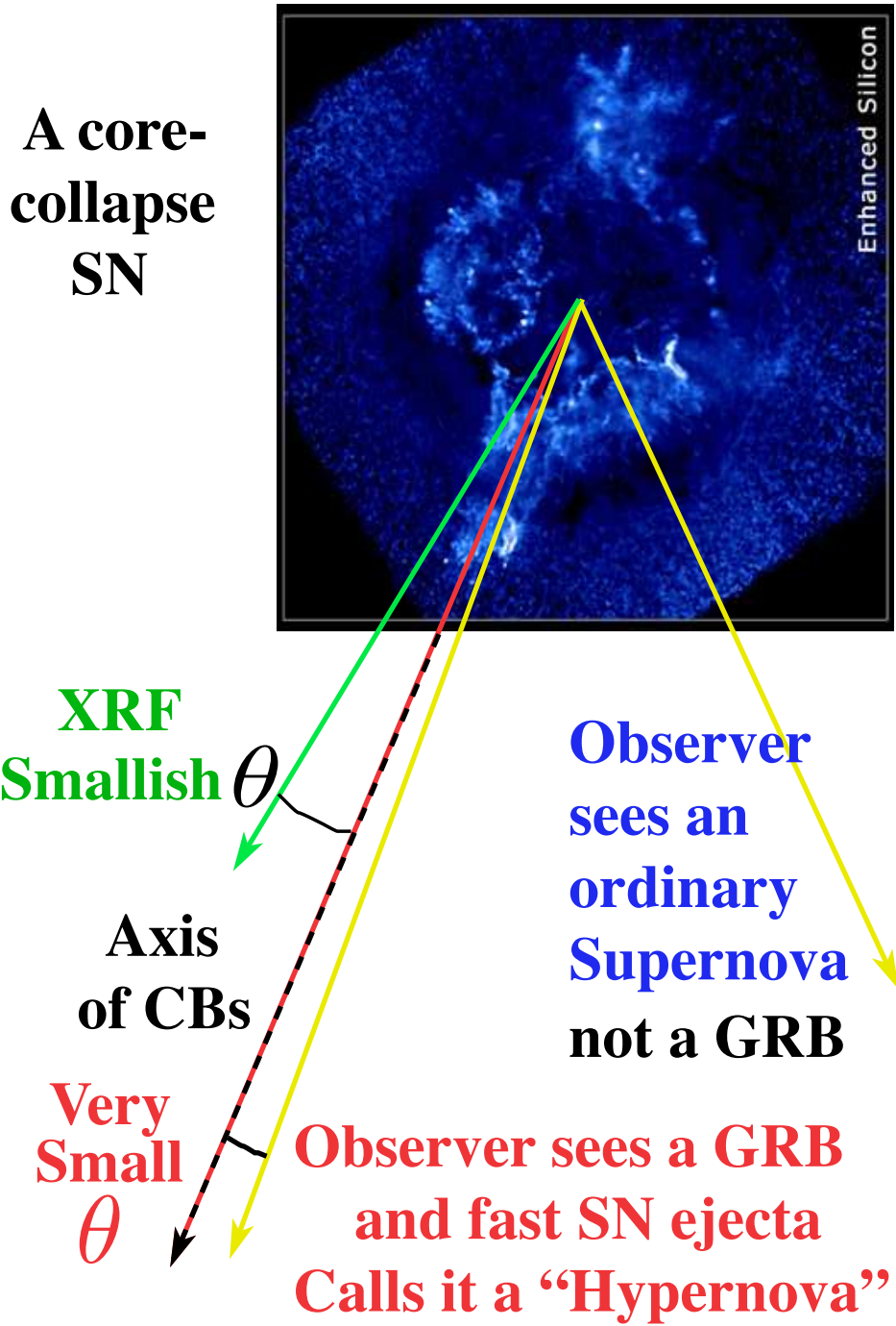,width=13cm}
\caption{Long GRBs and XRFs are seen close to the axis of their
associated supernova.
The nonrelativistic ejecta of a fairly ordinary SN may be
faster than average in its axial directions. Thus, {\it hypernovae}
(a hypothetical class of SNe associated with GRBs) may be 
in the eyes of the beholder. Cas A picture by NASA/CXC/GSFC/U.Hwang et al.}
\label{f2}
\end{figure}

In DDD02 and DDD03 we convinced ourselves that reality and the CB
model share some significant features. The hypotheses and physics
underlying the
prediction for the synchrotron-dominated phase were 
simple: equipartition of the magnetic field energy within a CB with that of 
the intercepted ISM particles, energy-momentum conservation for
the CBs' deceleration law, an initial relativistic expansion of a CB
in its rest system, followed by $F\!=\!m\,a$, for the evolution of
its radius. The resulting expressions provided good fits
to the broad-band AGs of {\it all} localized GRBs known at the time,
and allowed us to demonstrate, to our entire satisfaction, 
that a supernova (SN) akin to SN1998bw was seen in the data in
all cases were it could be observed, and unseen in all cases where
it couldn't (the distinction occurred at a redshift $z\!\sim\!1.2$).
This corroborated the model's original hypothesis (Dar \& Plaga 1999)
that core-collapse SNe are the engines of long GRBs, and gave
us the necessary confidence to predict the night in which SN2003dh,
associated with GRB030329, would be discovered (Dado et al.~2003b).
Since that night, everybody `had always known' that
the SN/GRB association was for real, see Fig.~\ref{f2}.

Given the information extracted from our AG analyses on the typical
Lorentz factors, observation angles and cannonball masses
of pre-SWIFT GRBs,
the observed properties of pre-SN winds, and the early luminosity of
a core-collapse SN (such as SN1987A),
we could, in DD04, go one step forward. The early light of a SN
scatters in its semi-transparent
`circumburst material', previously blown by the wind
of the progenitor. This creates a {\it glory}: a light-reservoir of 
non-radially directed photons. In DD04 we studied the outcome
of ICS of the light of the SN's
glory by the electrons enclosed in a CB.
The mechanism correctly predicts all the established
properties of GRB pulses, their typical values, and the distributions
around them. These include their peak-energies, 
equivalent isotropic energy, pulse shape, 
spectrum and temporal evolution. As Sidney Coleman would put it:
`Modesty restrains me, but honestly
compels me'... to admit that, as of then, I am convinced that
long GRBs have been understood. After all, even in astrophysics, a problem
is occasionally solved, e.g.~the energy source of stars.
GRBs are a small fraction of the energy budget of a supernova,
nothing like the advertised `biggest explosions after the Big Bang'.
Originally, the solutions to the grand problem of stars and the modest
problem of GRBs were incomplete:  Hans Bethe did not have a decent
theory of the weak interactions, nor do we have a predictive theory 
of the `cannon' (the accreting compact object resulting from a star's
core-collapse). In both cases, related observations filled these 
theoretical gaps.

\begin{figure}[]
\centering
\vbox{
\hbox{
 \psfig{file=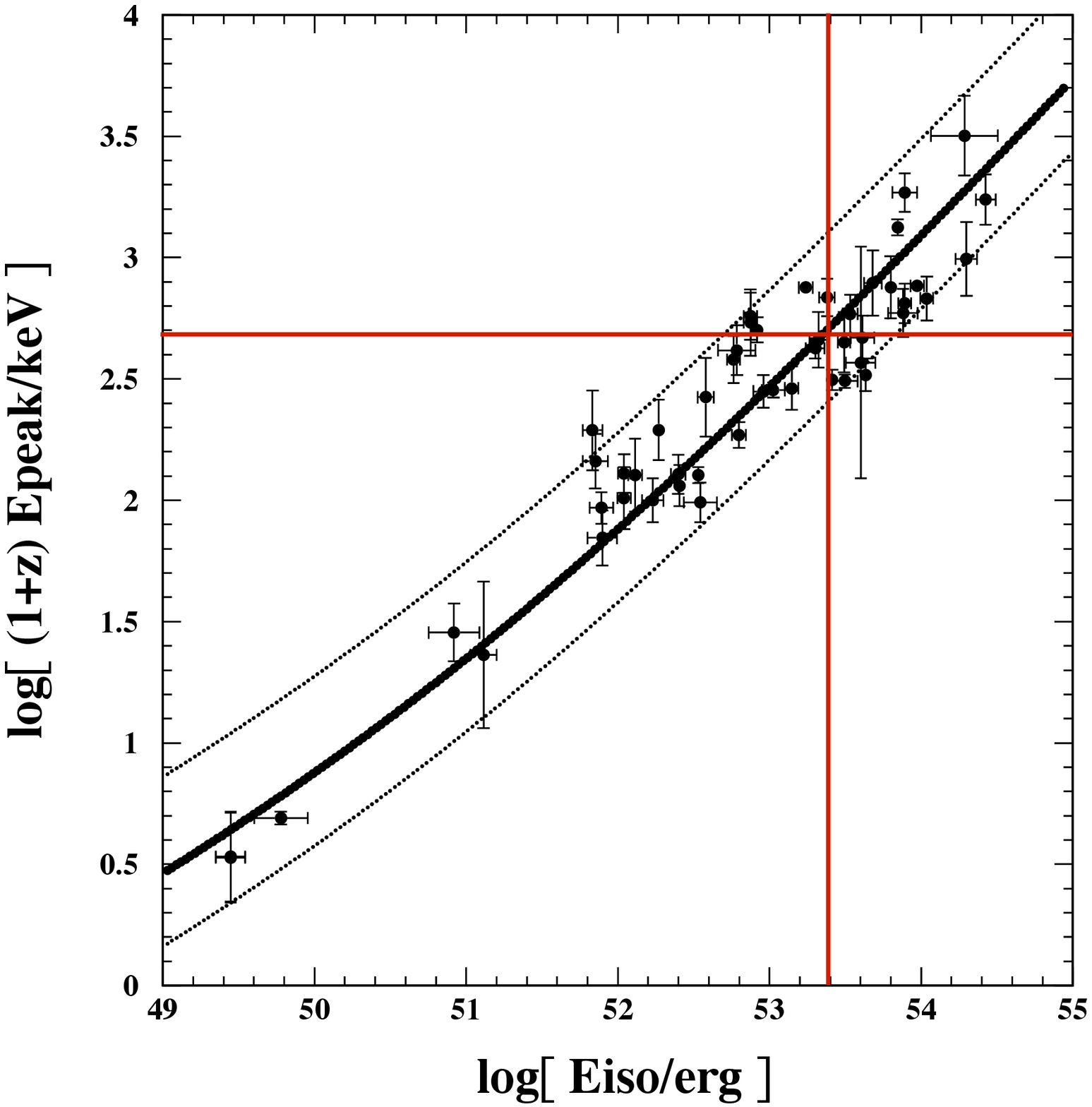,width=7.5cm}
 \psfig{file=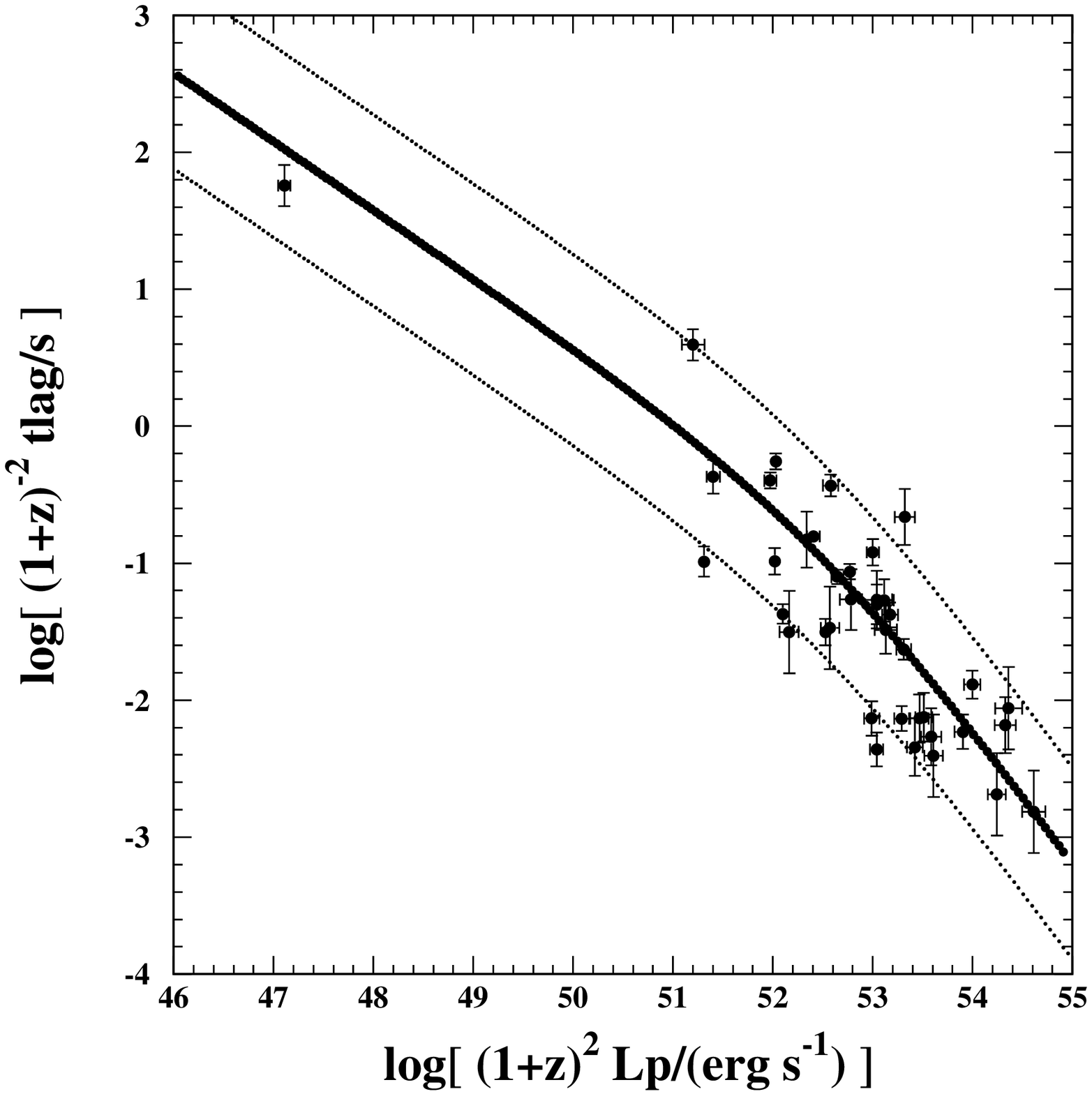,width=7.5cm }
}}
\centering
\vbox{
\hbox{
\psfig{file=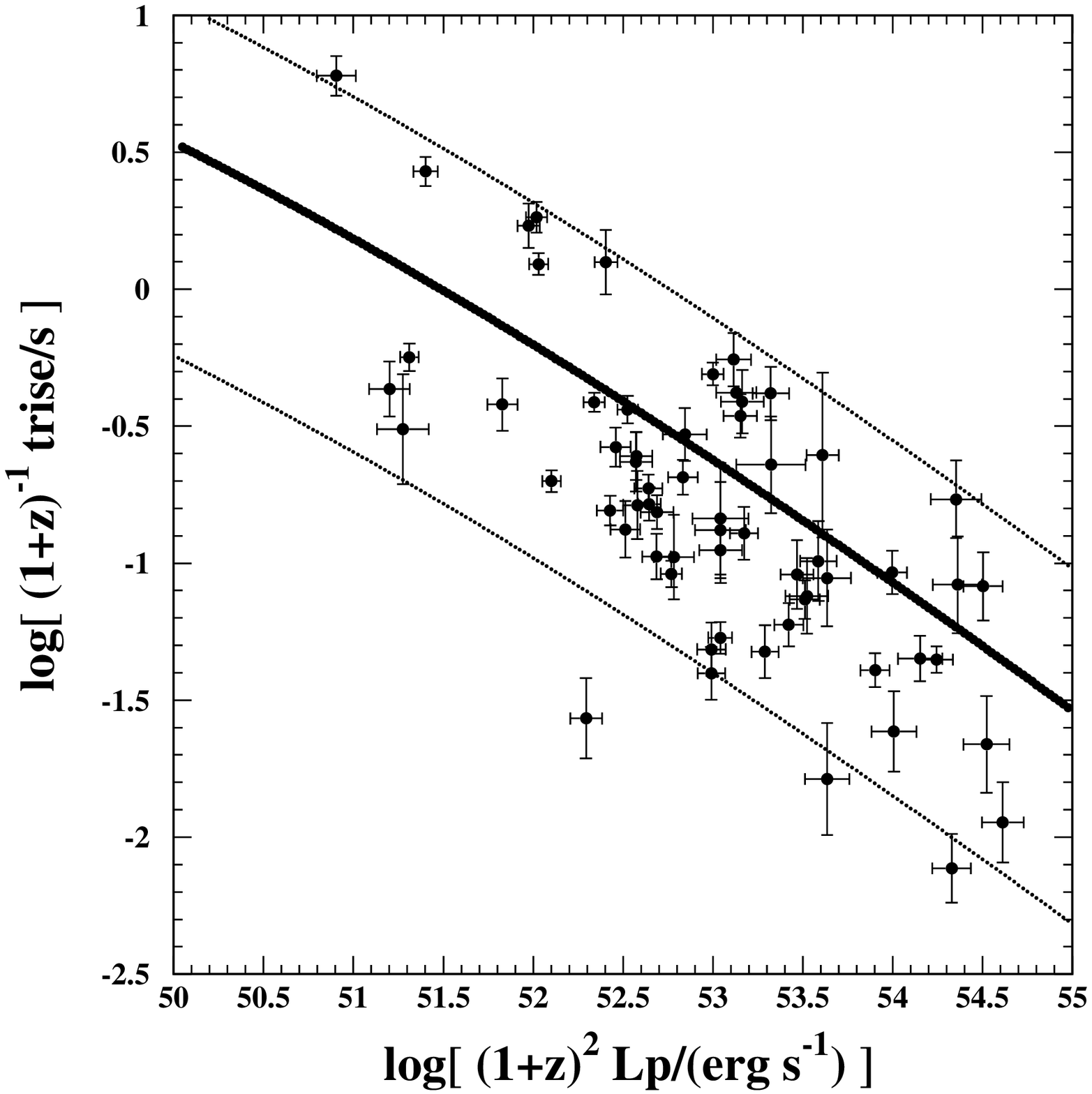,width=7.5cm}
\psfig{file=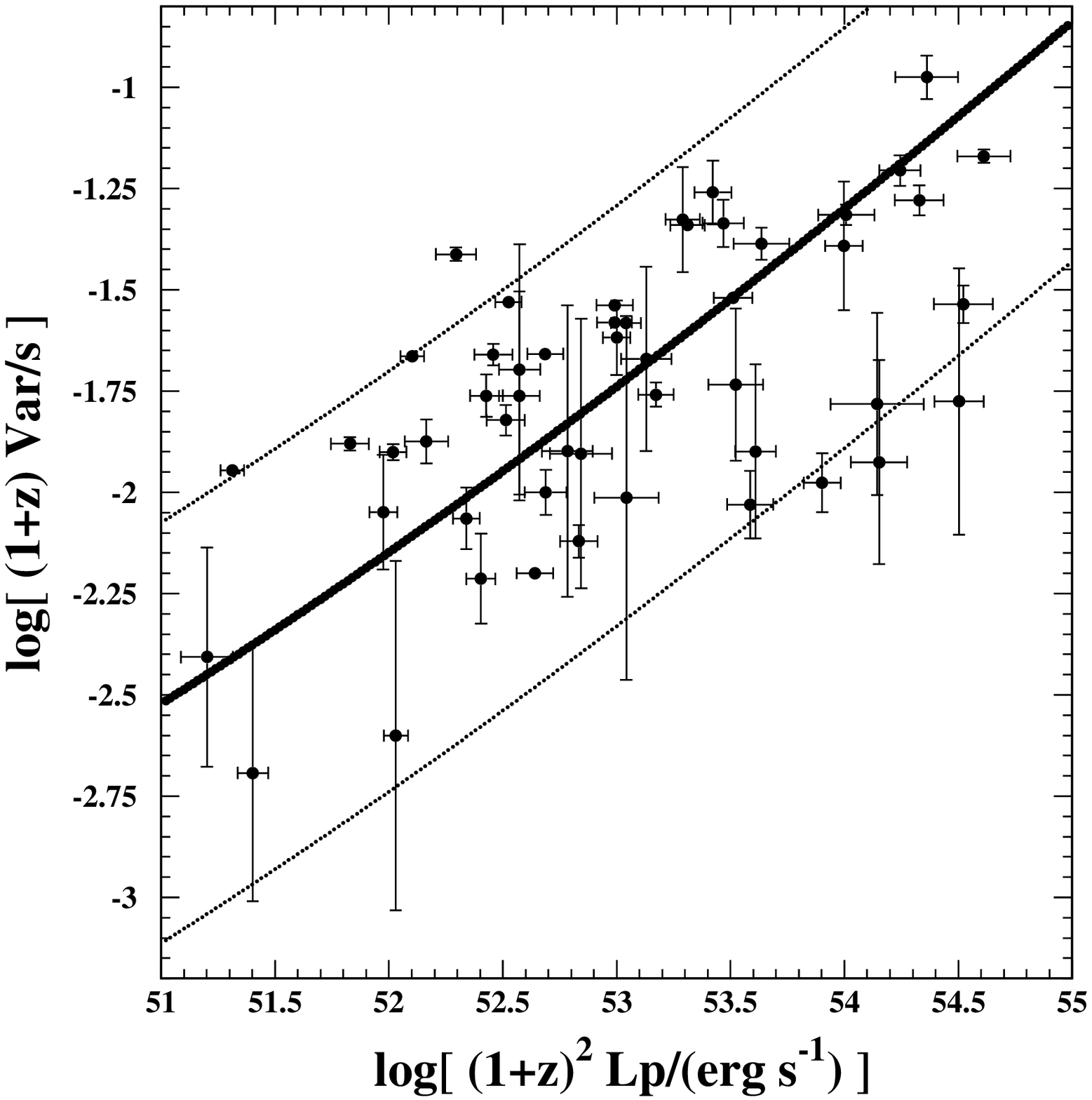,width=7.5cm}
}}
\vspace*{8pt}
\caption{{\bf (a) Top left}: The $(E_p,\,E_\gamma^{\rm iso})$ correlation, compared
with its predicted trend in the CB model. The crossed
red lines are the  predicted typical or average values.
{\bf (b) Top right}: The $[t_{\rm lag},L_p^{\rm iso}]$ correlation.
{\bf (c) Bottom left}: The $[t_{\rm rise},L_p^{\rm iso}]$ correlation.
{\bf (d) Bottom right}: The $[V,L_p^{\rm iso}]$ correlation. 
The definition and measurement of the rise-time of a pulse and,
more so, its variability, are debatable. Correlations involving these
quantities are the loosest. The $(E_p,\,E_\gamma^{\rm iso})$ correlation
compares two observables averaged over pulses (as opposed to 
an averaged observable and a single-pulse one). It is
 expected and observed to be the tightest: strictly speaking
Eqs.~(\ref{brief}) are for single pulses (Dado et al.~2007a).}
\label{f3}
\end{figure}

 Most of the rest of this note is devoted to how precisely ICS describes
the early X-ray and optical data. As an appetizer, I shall demonstrate
 that ICS is indeed the prompt
mechanism of GRBs and XRFs. Many correlations 
between GRB observables have been studied with the help of the 
new data on GRBs of measured $z$ (e.g.~Schaefer~2006).  
 All of the successful correlations are appallingly 
 simple predictions of the CB model:

The typical Lorentz factor of the CBs of GRBs is
$\gamma\!=\!{\cal{O}}(10^3)$. 
Expanding in its rest system at
a speed comparable to that of light, a CB is 
quasi-point-like: the angle it subtends from its point
of emission (in the SN frame) is  comparable or
smaller than the opening angle, 
$1/\gamma$, characterizing its beamed radiation.
Let  $\theta$ be  the viewing angle of an 
observer of a CB, relative to its direction of motion,
 typically of ${\cal{O}}$(1 mrad), and let $\delta$
be the corresponding Doppler factor:
$\delta\!\equiv\!{1/[\gamma\,(1-\beta\, \cos\theta)]}
 \! \simeq \! {2\, \gamma/[1+\gamma^2\, \theta^2]}$.
 To correlate two GRB observables, all one needs to know is
their functional dependence on $\delta$ and $\gamma$. This is because 
the $\theta$ dependence of $\delta(\gamma,\theta)$  is so pronounced, that 
it should be the
largest source of the case-to-case spread in the measured quantities. 
In the CB model,
the $(\gamma,\delta,z)$ dependences of the  
 spherical equivalent energy of a GRB, $E_\gamma^{\rm iso}$; 
 its peak isotropic luminosity $L_p^{\rm iso}$;
its peak energy, $E_p$ (Dar \& De R\'ujula 2001a);
its pulse rise-time $t_{\rm rise}$; 
its variability, $V$; and its
`lag-time', $t_{\rm lag}$ (Dado et al.~2007a); are\footnote{The coefficients of
proportionality in Eqs.~(\ref{brief}) have explicit dependences on 
the number of CBs in a GRB,  their
initial expansion velocity and baryon number.
With typical values fixed by the analysis of GRB afterglows,
the predictions agree with the
observations, see DD04, Dado et al.~(2007a),
and the crossed lines in Fig.~\ref{f3}a.}:
\begin{equation}
E_\gamma^{\rm iso}\propto\delta^3;\;\;
L_p^{\rm iso}\propto {\delta^4\over (1+z)^2};\;\;
E_p\propto {\gamma\,\delta\over 1+z};\;\;
{ t_{\rm rise} }\propto {1+z\over \gamma\,\delta};\;\;
V\propto  {\gamma\,\delta\over 1+z};\;\;
t_{\rm lag}\propto {(1+z)^2\over \delta^2\,\gamma^2}.
\label{brief}
\end{equation}
One obvious consequence is 
$E_\gamma^{\rm iso}\!\propto\![(1+z)^2\,L_p^{\rm iso}]^{3/4}$,
see Fig.~\ref{f4}a. 
The most  celebrated correlation is the $[E_p\,,E_\gamma^{\rm iso}]$ 
one, shown in Fig.~\ref{f3}a as a test of our prediction 
(Dado at al.~2007 and references therein). It evolves from
$E_p\!\propto\![E_\gamma^{\rm iso}]^{1/3}$ for small $E_p$, to  
$E_p\!\propto\![E_\gamma^{\rm iso}]^{2/3}$ for large $E_p$.
This is because for $\theta\!\ll\! 1/\gamma $, $\delta\!\propto\!\gamma$,
while in the opposite case $\delta$ and $\gamma$ are independent.
The correlations are
specific to the ICS by the CB's electrons
(comoving with it with a Lorentz factor $\gamma$) of the glory's photons,
that are approximately isotropic in the supernova rest system, and
are Doppler-shifted by the CB's motion by a factor $\delta$ 
(the result would be different, for instance, for SR
from the GRB's source, or self-Compton scattering of photons
comoving with it). Five correlations 
between the six observables in Eq.~(\ref{brief}) are independant,
three more, shown in Figs.~\ref{f3}b,c,d, complete a set.
The data agree with the predicted correlations.
QED.

 Another tell-tale signal of ICS is the polarization, predicted to be
$\Pi\!=\!2\,\theta^2\,\gamma^2/(1 + \theta^4\,\gamma^4)$ (Shaviv \& Dar
1995). With limited sensitivity, the most probable observation angle 
of a GRB is 
$\theta\!\sim\! {\cal{O}}(1/\gamma)$, resulting in $\Pi\!\sim\!1$.
In the CB model XRFs and long GRBs are the 
same objects, viewed at different $\theta$.
The former (conventionally defined by their low $E_p$)
are observed at larger $\theta$ than the latter (DD04). Thus, for
 XRFs, $\Pi$ is predicted to be smaller. Four GRB polarization
measurements have been reported, all compatible with $\Pi\!=\!100$\%.
But the data have large statistical and
(possibly) systematic errors (Coburn and Boggs, 2003; Willis et al.~2005;
Kalemci et al~2006;
Dado et al. 2007b, and references therein).

\section{More recent and detailed results}

\begin{figure}[]
\vspace{-1cm}
\centering
\vbox{
\hbox{
 \psfig{file=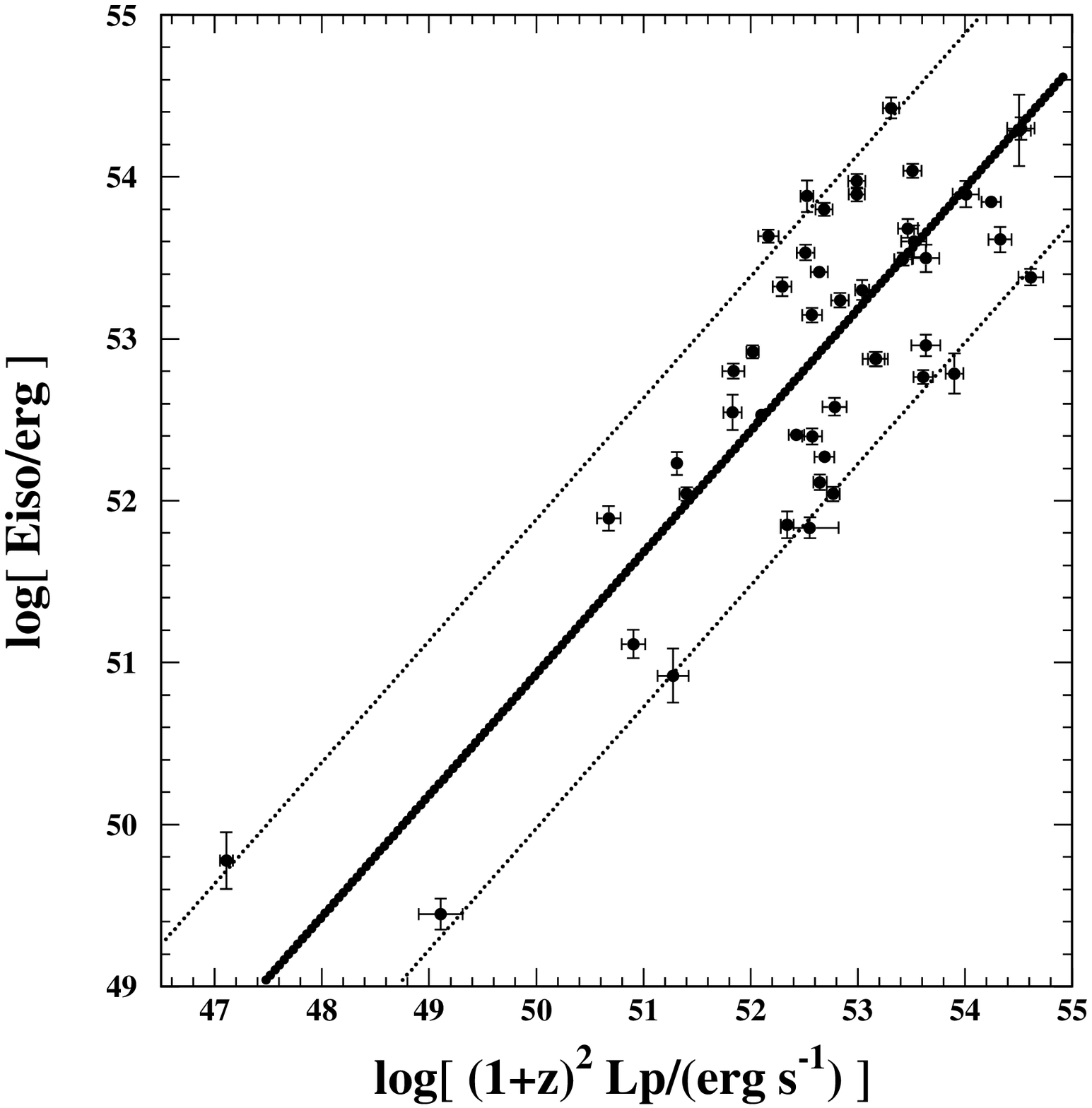,width=7.5cm}
 \hskip -.75cm
 \psfig{file=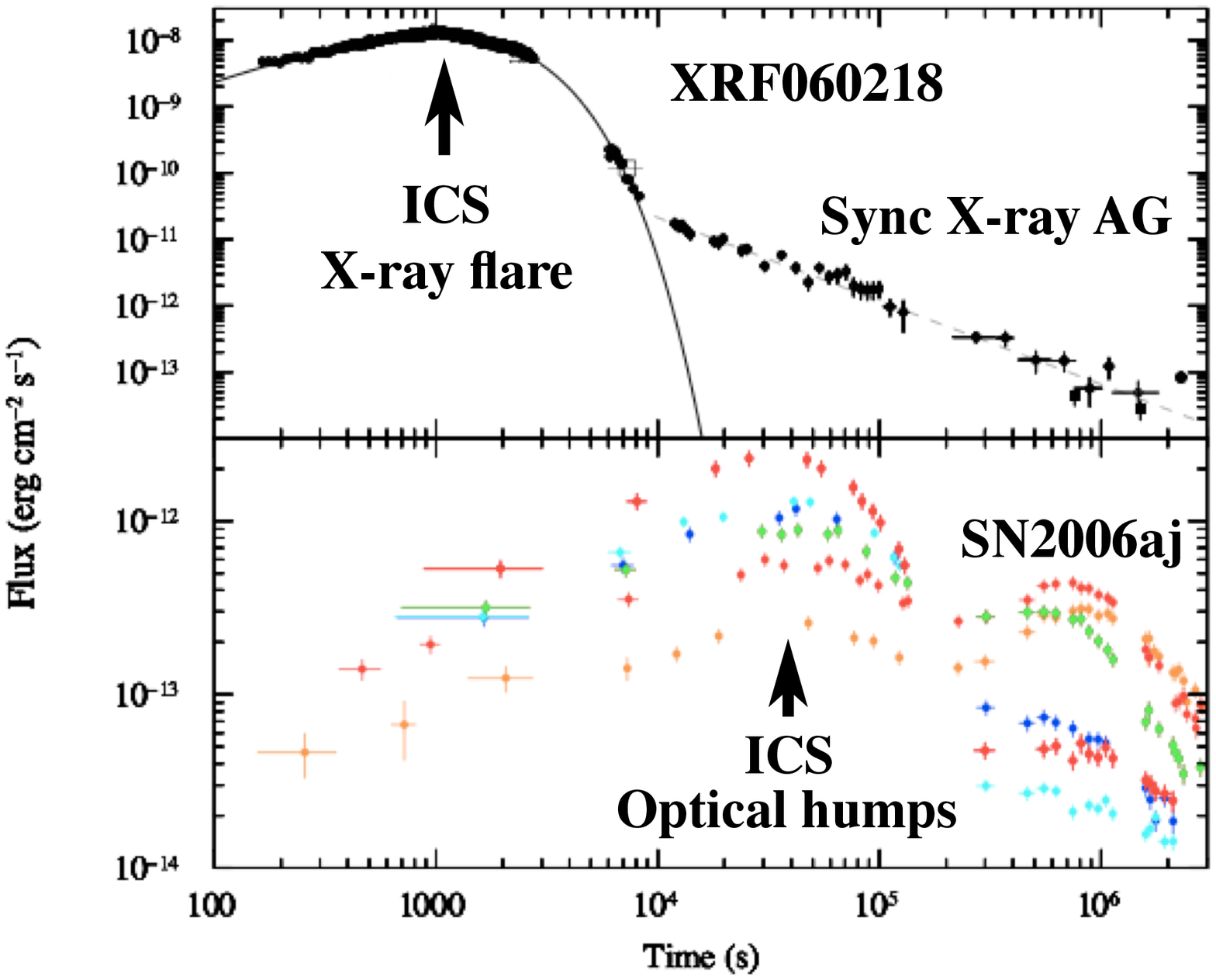,width=9.5cm }
}}
\centering
\vbox{
\hbox{
\psfig{file=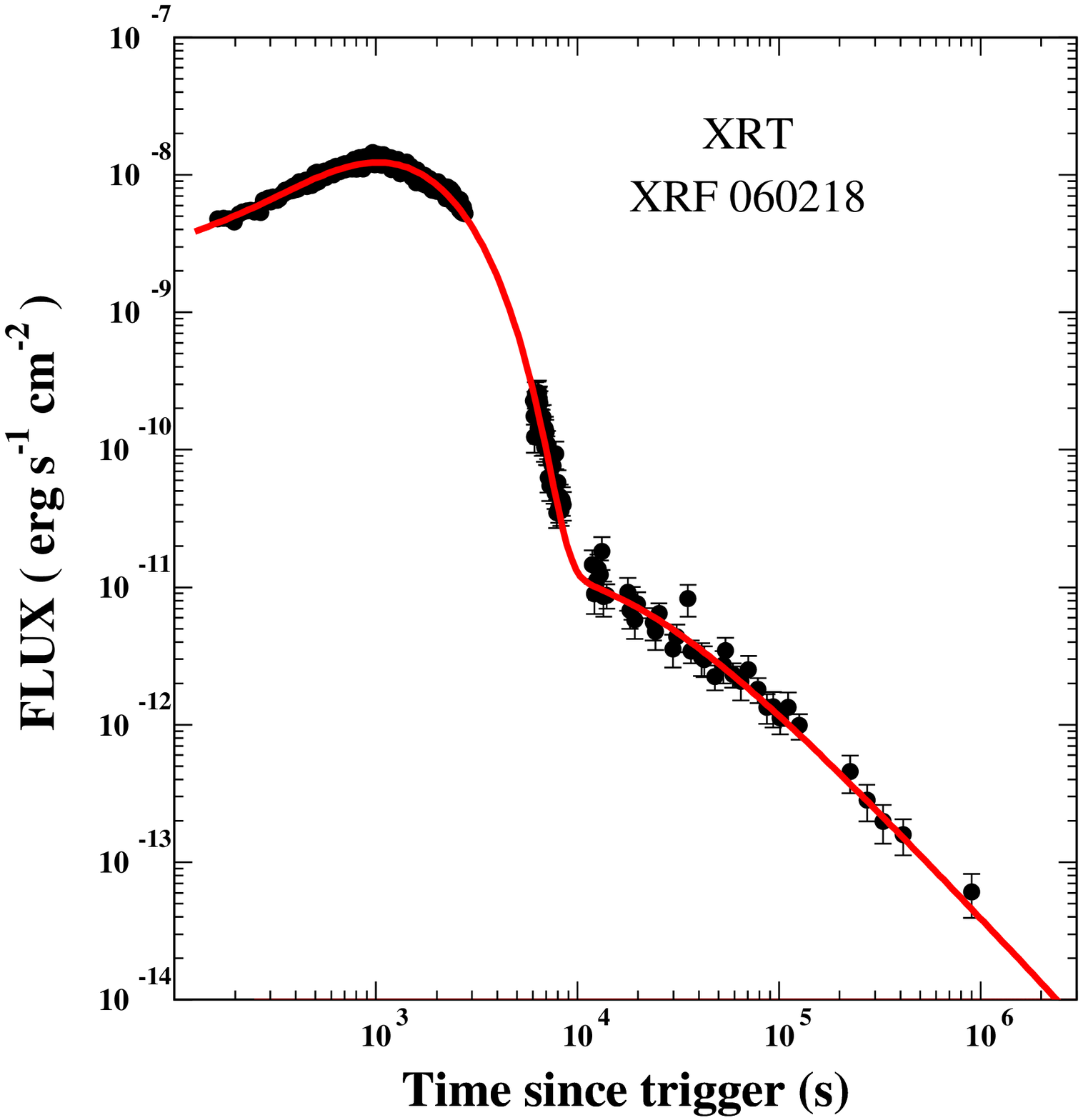,width=7.5cm}
\psfig{file=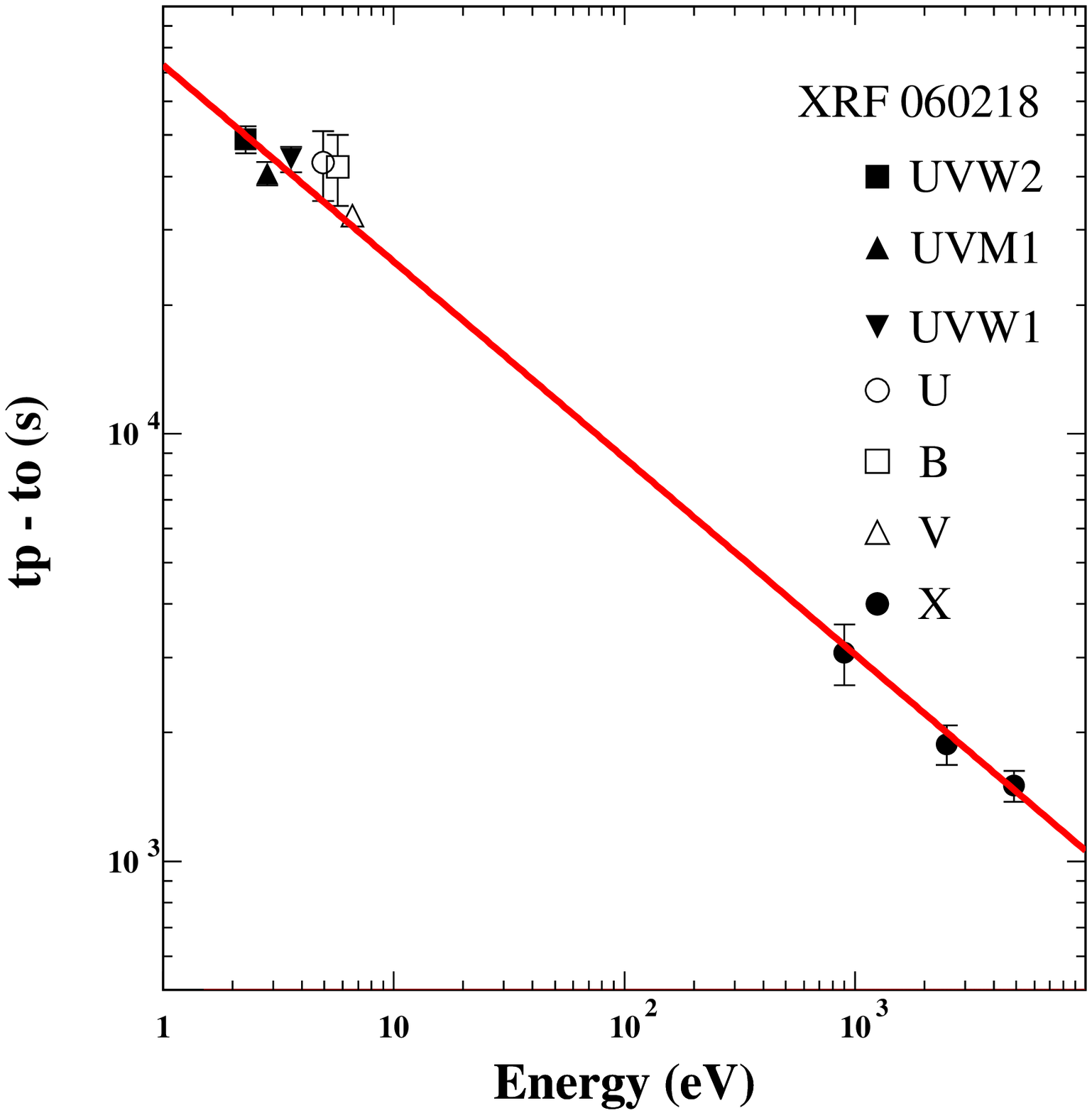,width=7.5cm}
}}
\vspace*{8pt}
\caption{{\bf (a) Top left}: 
The $[E_\gamma^{\rm iso},L_p^{\rm iso}]$ correlation (Dado et al.~2007a).
{\bf (b) Top right}: 
Data on XRF 060218/SN2006aj. Upper part:
the 0.3-10 keV SWIFT-XRT light curve, with
fits by Campana et al.~(2006b). Lower part: UVO light curves.
The ICS X-flare and UV/optical `humps' are our interpretation
(Dado et al.~2007c). 
{\bf (c) Bottom left}: The X-ray light curve of XRF 060218, fit in the
CB model.
{\bf (d) Bottom right}: Results of the `$E\!\times\!t^2$ law' for this XRF
 (Dado et al.~2007c).}
\label{f4}
\end{figure}

The XRF060218/SN2006aj pair provides one of the best testing 
grounds of theories, given its proximity ($z\!=\!0.033$), 
its excellent sampling and statistics, and its
simplicity: it had just one X-ray flare. The X-ray and optical
data, shown in Fig.~\ref{f4}b, have been modeled 
by a sum of a black body emission with a 
time-declining temperature and a cut-off power-law, allegedly
the result of the core-collapse shock breaking out from the 
stellar envelope and of the stellar wind of the SN's progenitor 
(Campana et al.~2006; Blustin 2007; Liang et al.~2007; Waxman, Meszaros \& 
Campana~2007). No novelties are required to 
understand this XRF/SN pair in the CB model. I use it next as an exemplar
to introduce and test some very specific predictions.

A typical CB has 
$E_\gamma^{\rm iso}\!\approx\! 0.8 \times 10^{44}\,\delta^3$ erg
and its {\it early} $E_p$ is 
$E_p(0)\!\approx\! 2\,\gamma\,\delta\, T_g/[3(1+z)]$,
with $T_g\!\sim\! 1$ eV the (pseudo)-temperature of the glory's
 thin-bresstrahlung spectrum 
$E_i\,dN_i/dE_i\!\sim\! {\rm exp}[-E_i/T_g]$ (DD04). 
Thus, the reported $E_{\rm iso}\approx (6.2\pm 0.3)\times 10^{49}$ erg 
and $E_p(0)\!=\!54$ keV result in the estimates
$\delta_0\!\sim\! 92$, $\gamma_0\!\sim\! 910$
and $\theta\!\sim\! 4.8\times 10^{-3}$.
Using these parameters and a fit value for the ISM density
we can predict the SR contribution to the X-ray AG
in the 0.3-10 keV band. It is shown in Fig.~\ref{f4}c, it dominates
after $t\!\sim\!10^4$ s. The shape of the initial ICS-dominated
X-ray peak is a prediction, its normalization and width 
are fitted, but they are totally compatible with the expectations
for the above $\gamma$ and $\theta$ values  (Dado et al.~2007c). 

A quick look at the `prompt' $\gamma$-ray count-rates and X-ray
light curves of SWIFT GRBs reveals that the $\gamma$-ray `pulses'
and X-ray `flares' are coincident. In the CB-model
this relation is very well understood. The derived ICS
spectrum bears a striking resemblance to the observed
`Band' spectrum. A spectrum  $E\,dN/dE\!\sim\! {\rm exp}[-E/E_p]$
generally suffices to describe the SWIFT X-ray data 
(that extend up to 350 keV). Such a spectrum is
simply the one generated by electrons comoving
with a CB, Compton up-scattering the glory's photons.
In more detail, a single GRB pulse or X-ray 
flare obeys a `master formula' (DD04, Dado et al.~2007c):
\begin{eqnarray}
E\, {d^2N_\gamma\over dt\,dE}&\approx& F[E\,t^2]\approx
 \Theta[t]\;
e^{-[\Delta t(E)/t]^2}\, 
 \left\{1-e^{-[\Delta t(E)/t]^2}\right\}\, e^{-E/E_p(t)},\nonumber\\
 E_p(t)&\approx& E_p(0)\left( 1-{t/\sqrt{\Delta t^2+t^2}}\right),
\;\;\;
 \Delta t(E)\approx t_{tr}^w \left(E_p/E\right)^{1\over 2},
 \label{master} 
\end{eqnarray}
where $t_{tr}^w$ is the observer's time at which the remaining
`wind' material becomes transparent, of ${\cal{O}}(1\rm s)$, for typical
parameters. The energy fluence $F$ of Eq.~(\ref{master}) is predicted
to be approximately a function of the combination $E\!\times\! t^2$ of its
two variables (DD04). This is mainly due to the fact
that a glory's photon incident on a CB
(in the supernova rest system) at an angle $\alpha$ is Compton
scattered to an energy 
$E\!\propto\!1+\cos\alpha\!\to\! 1/r^2\!\propto\!1/t^2$. The
limit is for the large times, $t\!\gg\! t_{tr}^w$ 
[or radial distances, $r\!\gg\! c\,\gamma\,\delta\, t_{tr}^w/(1+z)$]
for which the glory's light appears to the escaping CB to be radially 
distributed. But even at $t\!\sim\! t_{tr}^w$, by definition, the
wind material is semitransparent and the distribution of
the light it scatters is such that $\langle\cos\alpha\rangle\approx -1+ 1/r^2$,
making the $E\!\times\!t^2$ `law' a good approximation at all $t$.
The predicted $E_p(t)$ evolution 
is observed in the time-resolved spectra of well isolated pulses 
(see, for instance, the insert in Fig.~8 of Mangano et al.~2007).

For XRF060218,
Eq.~(\ref{master}) describes very well the flare and the
fast-decay of the X-ray light curve, see Fig.~\ref{f4}c,
and the fast softening of its spectrum
during this phase. As soon as  SR dominates over ICS 
(at $t\!\sim\! 9000$ s) 
the spectrum 
becomes a harder power-law with an observed index
$\beta_X\sim -1.1$, the prediction of the CB model for the 
unabsorbed synchrotron spectral energy density in the  X-ray band 
(DDD02).

The most impressive CB-model result concerning XRF060218
is the prediction of the properties of its optical emission. We contend that
the `optical humps' in Fig.~\ref{f2}c are nothing but the X-ray flare,
seen at much lower frequencies. This we prove by fitting the 
15-150 keV band data to Eq.~(\ref{master}) to extract the
peak time, $t_{\rm peak}(E_{\rm X})\!=\! 425\pm 25$ s, and 
peak energy flux,
$\rm PEF\!=\!1.30\pm 0.07\,\rm erg\,cm^{-2}\,s^{-1}$. The peak times
in 2 other X-ray energy intervals and in 6 UV and optical
frequencies are then predicted: 
$t_{\rm peak}(E_j)\!\approx\!t_{\rm peak}(E_{\rm X})\sqrt{E_j/E_{\rm X}}$,
$j\!=\,1,8$. The results are shown in Fig.~\ref{f4}d, an extension
by $>$ 3 orders of magnitude of similar results in DD04. The 
(spectral) results for the PEFs are equally
simple and successful (Dado et al.~2007c). 
Recall that all these results follow from
the elementary physics of ICS on the glory's light of a supernova.
Finally, this XRF is indeed associated
with a SN, since it is simply a (long) GRB seen
at a relatively large $\theta$. Sometimes a single
observation, e.g.~the $\Omega^-$, or data set, e.g.~the spectrum
of charmonium, suffices to definitely
establish a theory. The XRF060218/SN2006aj
pair would be such a case... if astrophysics was more akin to
particle physics, or to today's cosmology.

\begin{figure}[]
\centering
\vbox{
\hbox{
 \psfig{file=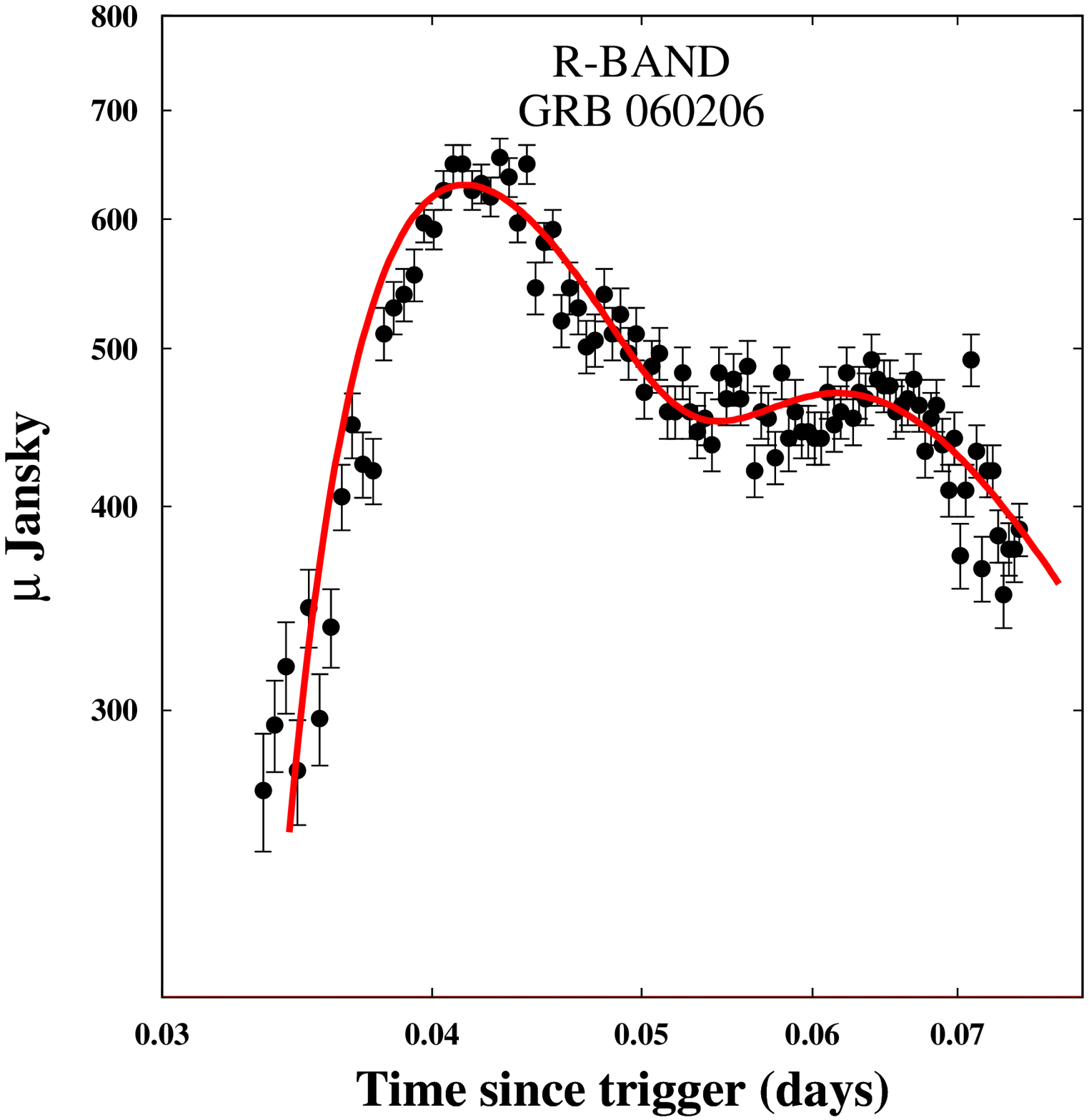,width=7.5cm }
 \psfig{file=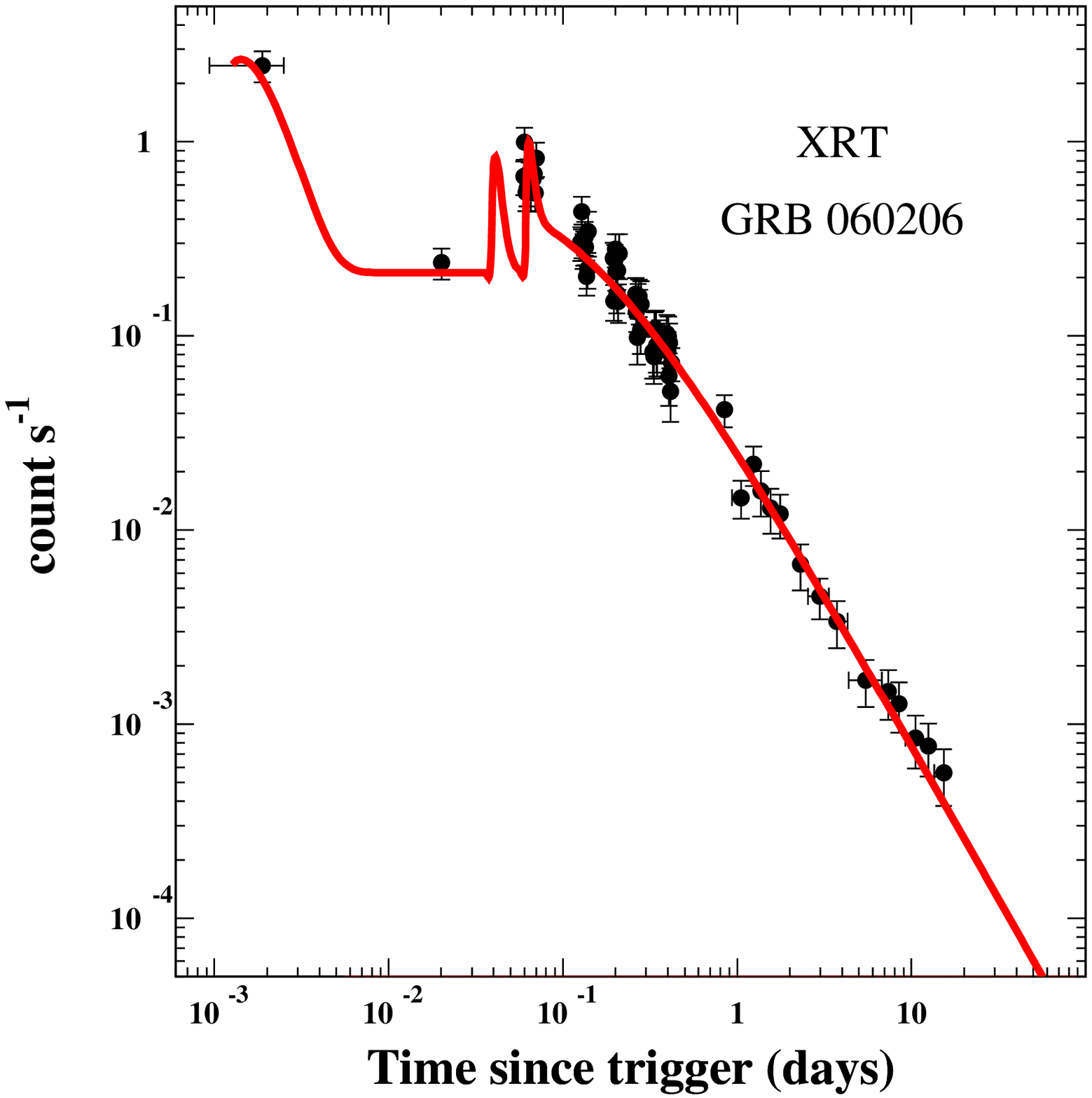,width=7.5cm}
}}
\centering
\vbox{
\hbox{
\psfig{file=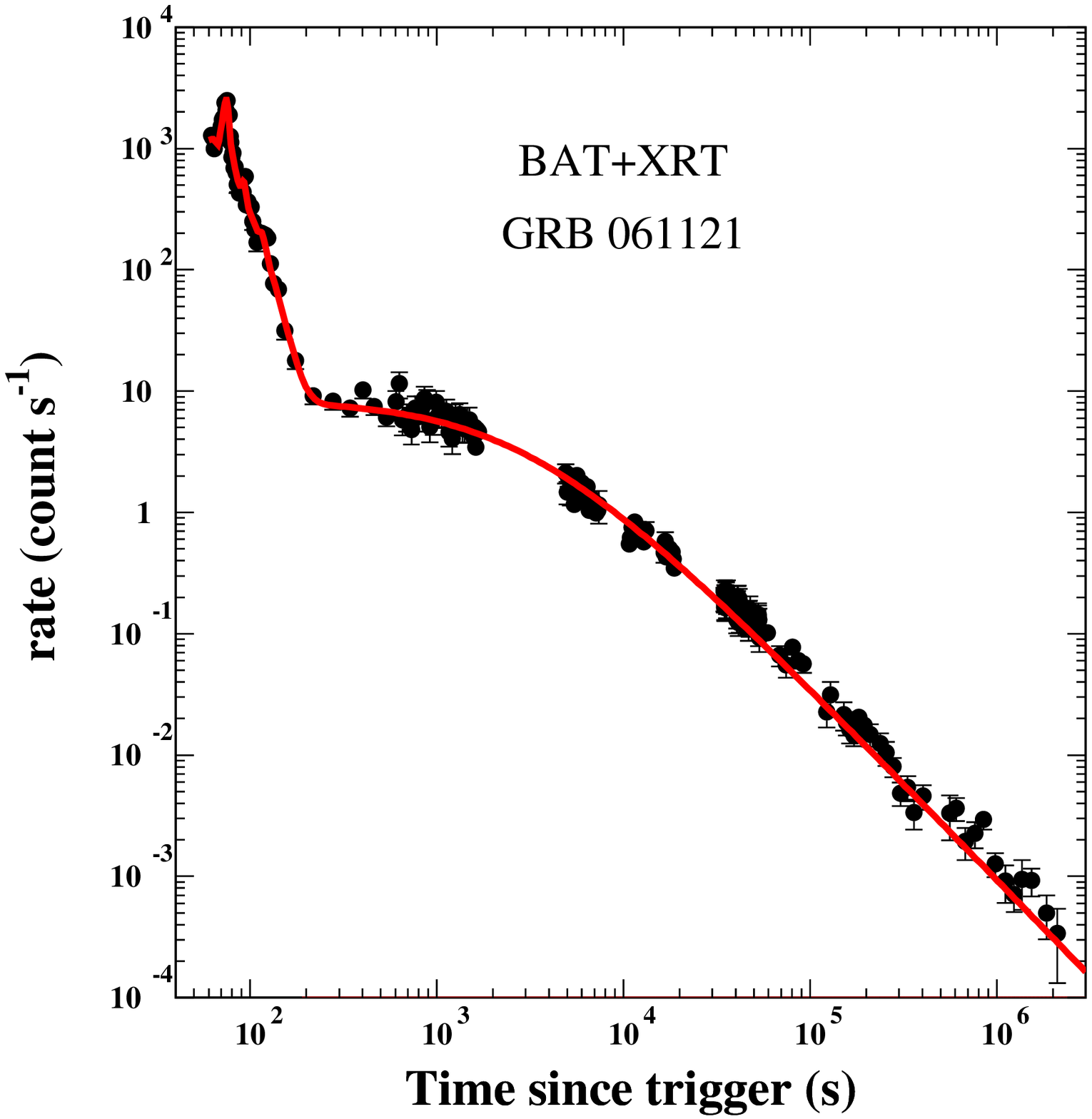,width=7.5cm}
\psfig{file=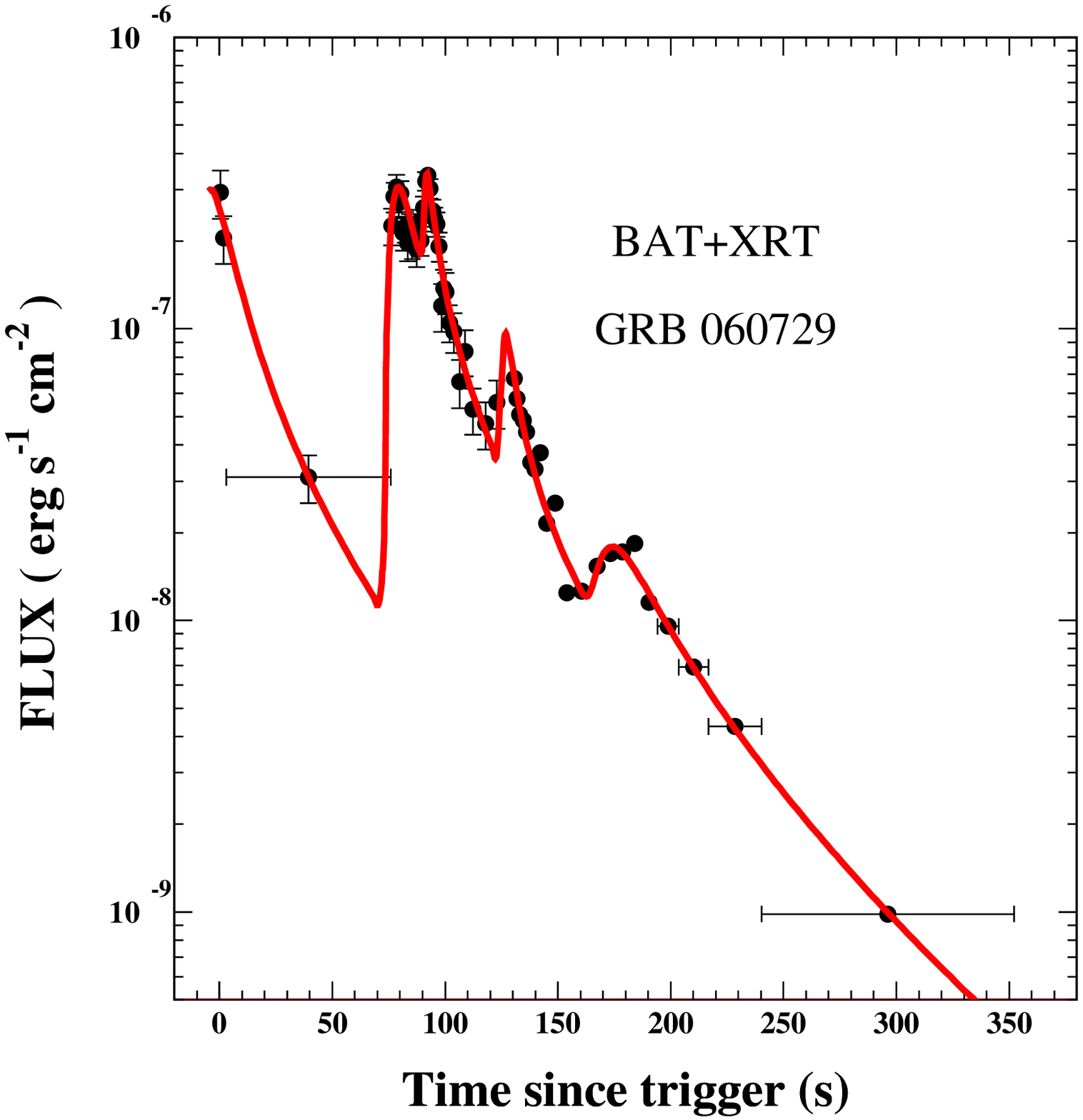,width=7.5cm}
}}
\vspace*{8pt}
\caption{{\bf (a) Top left}: The very early
R-band light curves of GRB 060206 (Wozniak et al.~2006).
{\bf (b) Top right}: The less-well sampled X-ray light curves 
(Evans et al.~2007),
inferred from  SWIFT BAT and XRT data for GRB 060206. 
The properties of the last two prominent X-ray flares are
predicted from those of their optical counterparts in {\bf{a}}.
{\bf (c) Bottom left}: The extensive X-ray light
curve of GRB 061121 (Page et al.~2007). 
{\bf (d) Bottom right}: The complex decay of the X-ray light curve
of GRB 060729 (Grupe et al.~2007), with the progressively
dying pangs of its accreting engine. 
All data are accompanied by their CB-model description
(Dado et al.~2007c).}
\label{f5}
\end{figure}

In Dado et al.~(2007c) we have analized a sample of GRBs and XRFs
which includes the brightest 
SWIFT GRBs, the one with the longest measured X-ray emission, a couple 
with canonical X-ray light curves (with and without late X-ray `mini-flares') 
and some 
of the ones considered most puzzling from the point of view of fireball 
models. In Fig.~\ref{f5} I show some of the results. In the case of
GRB060206, the very early optical data, shown in Fig.~\ref{f5}a,
are exceptionally good. We fit it, and use Eqs.~(\ref{master}) to 
predict the X-ray light curve from $t\!\sim\!0.03$ s onwards,
see Fig.~\ref{f5}b.
The complete X-ray light curve of GRB061121, was measured over 
a record seven
 orders of magnitude in intensity and five orders of magnitude in time.
 It is impressively well described by the simple predictions
 of the CB model for ICS followed by SR, as shown in Fig.~\ref{f5}c.
In some cases, during the rapidly
decreasing phase of the X-ray light curve, there are `mini-flares'
not seen in the corresponding $\gamma$-ray light curve.
One example is GRB060729, shown in Fig.~\ref{f5}d.
In the CB model, CBs are ejected
 in delayed accretion episodes  of matter from
a ring or torus (Dar and De R\'ujula, 2000). As the accretion
material is consumed, one may expect  the `engine' to have
a few progressively-weakening dying pangs, seen as miniflares.
We now know that, indeed, SNe `may bang more than once'
(De R\'ujula 1997). They do it at least as many times as we
see $\gamma$-ray peaks and X-ray flares in a GRB or an XRF.


\section{Conclusions}

The data on GRBs gathered after the launch of SWIFT, 
as interpreted in the CB model, 
has taught us four things:
\begin{itemize}
\item{}The relatively narrow pulses of the $\gamma$-ray signal, the
somewhat wider prompt flares of the X-rays, and the much wider humps
sometimes seen at UVO frequencies, have a common origin. They are
generated by inverse Compton scattering.
\item{} Relatively weak `dying pang' episodes of accretion are seen as X-ray
`mini-flares' in the declining phase of the previous mightier pulses,
which are seen both in $\gamma$ and X-rays.
\item{} The historical distinction between prompt and afterglow phases
is obsolete. It is replaced
by a physical distinction: the relative dominance of the Compton or
synchrotron mechanisms at different times, which are
frequency-dependent.
\item{} The two quoted mechanisms 
suffice to provide a very simple and accurate description
of XRFs and long-duration GRBs at all frequencies and times.
They generate the rich structure
of the light curves at all frequencies, and their chromatic or
achromatic `breaks'.
\item{} To date, as the quality of the data improves, so does the 
 quality of its agreement with the CB-model's predictions.
\end{itemize}
The above `SWIFT teachings' are not the ones most 
specialists in the field would list.
 The fast decline,  `continued engine activity' and misplaced `break'
of the `canonical' behaviour are no doubt the most important 
recognized new challenges to the orthodox --but ductile-- views. 
In the CB model they are 
predictions. Some other SWIFT teachings
are also indisputable, e.g.~the average
$z$ of SWIFT GRBs is 2.8, and its current record is 6.29.

When their collimated radiation points to the observer, GRBs 
are the brightest sources in the sky.
In the simple-physics context of the CB model, 
GRBs are not  persistent mysteries,
and a constant source of surprises, exceptions and new requirements. 
Instead, they are well-understood and can be used as cosmological tools, 
to study the history of the intergalactic medium and of star formation up to 
large redshifts, and to locate SN explosions at an  early stage.
In the CB model GRBs are not `standard candles',
their use in `Hubble-like' analyses would require further
elaboration. The GRB conundra have been reduced to just one:
`how does a SN manage to sprout mighty jets?'
The increasingly well-studied CBs ejected by quasars and microquasars,
no doubt also fired in 
catastrophic accretion episodes on compact central objects, 
provide observational hints with which, so far, theory
and simulations cannot compete.

The CB model underlies a unified theory of high energy astrophysics.
The information gathered in our study of GRBs can be
used to understand, also in very simple terms, other phenomena.
The most notable is (non-solar) Cosmic Rays. We allege  (Dar 
et al.~1992; Dar \& Plaga~1999)  
that they are simply the charged ISM particles scattered by 
CBs, in complete analogy with the ICS of light by the same CBs.
This results  in a successful
description of the spectra of all primary cosmic-ray
nuclei and electrons at all observed energies
(Dar and De R\'ujula~2006a). The CB model also predicts very
simply the spectrum of the gamma background radiation 
and explains its directional properties (Dar \& De R\'ujula~2001b; 2006b).
Other phenomena understood in simple terms include the
properties of cooling core 
clusters (Colafrancesco, Dar \& De R\'ujula~2003) and 
of intergalactic magnetic fields 
(Dar \& De R\'ujula~2005). The model may even have a say in
`astrobiology' (Dar et al.~1998; Dar \& De R\'ujula~2001b).

Finally, if the cannonballs of GRBs are so pervasive, one may ponder
why they have not been directly seen. After all, particularly in 
astrophysics, {\it seeing is believing}. The answer is simple.
Cannonballs are tiny astrophysical objects: their typical mass is 
half of the mass of Mercury.
Their energy flux at all frequencies is $\propto\!\delta^3$,
large only when their
Lorentz factors are large. But then, the radiation is also
extraordinarily collimated, it can only be seen nearly on-axis.
Typically, observed SNe are too far to {\it photograph} their CBs
with sufficient resolution.

Only in two SN explosions that took place close enough, the
CBs were in practice observable.  One case
was SN1987A, located in the LMC,
whose approaching and receding CBs were
photographed by Nisenson and Papaliolios (2001). 
The other case was SN2003dh, associated with GRB030329,
at $z=0.1685$. In the CB model interpretation,
its two approaching CBs were first `seen'
as the two-peak $\gamma$-ray light curve and the two-shoulder
AG (Dado et al.~2003b). This allowed us 
to estimate the time-varying angle of their superluminal
motion in the sky. Two sources or `components'
were indeed clearly seen in radio observations
at a certain date, coincident
with an optical AG rebrightening. We claim
that the data agree with our expectations\footnote{The
size of a CB is small enough to expect its radio image to
scintillate, arguably more than observed (Taylor et al.~2004).
We only realized a posteriori that the ISM electrons a CB
scatters, synchrotron-radiating in the ambient magnetic field, would
significantly contribute at radio frequencies, somewhat blurring the 
CB's radio image (Dado, Dar \& De R\'ujula 2004,
Dado \& Dar 2005).},  including 
the predicted inter-CB superluminal separation 
(Dar \& De R\'ujula 2000, DD04).
The observers claimed the contrary, though the 
evidence for the weaker `second component' is $>20\sigma$,
and it is `not expected in the standard model' (Taylor et al.~2004).
The no-doubt spectacular-discovery
picture of the two superluminally moving sources has, to my
knowledge, never been published. This is too bad, for {\it a picture is
worth a thousand words.}

More often than not, an insufficient but
 necessary condition for an astrophysical theory
to eventually survive as part of the truth is that it be long opposed, despised, 
and authoritatively
pronounced wrong by the community. The CB model passes this
final test with flying colours.

\begin{acknowledgements}
I am indebted to Shlomo Dado, Arnon Dar and Rainer Plaga for our
collaboration and to Franco Giovannelli for an excellent Frascati
Workshop 2007, at Vulcano.
\end{acknowledgements}




\end{document}